## Systems theory of Smad signalling

David C. Clarke[1,2], M. D. Betterton[3,4], and Xuedong Liu[1,4]

Departments of [1]Chemistry and Biochemistry, [2]Chemical and Biological Engineering, and [3]Physics

University of Colorado-Boulder, Boulder, Colorado, 80309

[4]Corresponding authors:      M. D. B.:      Tel: (303) 735-6135

                                              Fax: (303) 492-3352

                                              Email: mdb@colorado.edu

                              X. L.:          Tel: (303) 735-6161

                                              Fax: (303) 735-6161

                                              Email: xuedong.liu@colorado.edu

Running Title: Systems theory of Smad signalling

## Abstract

Transforming Growth Factor-β (TGFβ) signalling is an important regulator of cellular growth and differentiation. The principal intracellular mediators of TGFβ signalling are the Smad proteins, which upon TGFβ stimulation accumulate in the nucleus and regulate transcription of target genes. To investigate the mechanisms of Smad nuclear accumulation, we developed a simple mathematical model of canonical Smad signalling. The model was built using both published data and our experimentally determined cellular Smad concentrations (isoforms 2, 3, and 4). We found in mink lung epithelial cells that Smad2 (8.5-12 × 10$^4$ molecules/cell) was present in similar amounts to Smad4 (9.3-12 × 10$^4$ molecules/cell), while both were in excess of Smad3 (1.1-2.0 × 10$^4$ molecules/cell). Variation of the model parameters and statistical analysis showed that Smad nuclear accumulation is most sensitive to parameters affecting the rates of R-Smad phosphorylation and dephosphorylation and Smad complex formation/dissociation in the nucleus. Deleting Smad4 from the model revealed that rate-limiting phospho-R-Smad dephosphorylation could be an important mechanism for Smad nuclear accumulation. Furthermore, we observed that binding factors constitutively localised to the nucleus do not efficiently mediate Smad nuclear accumulation if dephosphorylation is rapid. We therefore conclude that an imbalance in the rates of R-Smad phosphorylation and dephosphorylation is likely an important mechanism of Smad nuclear accumulation during TGFβ signalling.

# 1. Introduction

Transforming growth factor-β (TGFβ) signalling controls diverse cellular processes, including cell proliferation, differentiation, migration, and apoptosis [1]. TGFβ is a member of the TGFβ superfamily of cytokines, which also includes bone morphogenic proteins, Mullerian inhibitory substance, activin, inhibin, and Nodal [2]. Defects in TGFβ signalling can lead to a variety of diseases, including cancer [3, 4]. Our ability to control TGFβ signalling during disease states depends on understanding the mechanisms of TGFβ signalling.

The principal components of TGFβ signalling are the active form of the TGFβ ligand, the membrane-bound TGFβ receptors, and the intracellular Smad proteins. The biologically active TGFβ ligand is a 25 kDa dimer connected by a disulfide bond formed between two cysteines on the monomeric units [5]. Ligand binding to the receptors located at the plasma membrane initiates signalling. Most TGFβ-sensitive cells express three distinct receptor proteins, the TGFβ type I, II, and III receptors [6]. The type I and type II receptors are serine/threonine kinases that transmit the TGFβ signal into the cell primarily via phosphorylation of the Smad proteins [7]. The eight predominant Smad isoforms are functionally classified into 3 groups: the receptor-regulated Smads (R-Smads; isoforms 1, 2, 3, 5, and 8), the common-mediator Smad (Co-Smad; isoform 4), and the inhibitory Smads (I-Smads; isoforms 6 and 7) [8]. The R-Smads are substrates for the receptors in the TGFβ family [2]. In this paper, we focus on Smad isoforms 2, 3, and 4 because these isoforms mediate TGFβ signalling [2, 9].

Canonical TGFβ signalling begins with exposure of cells to TGFβ ligand, which leads to receptor activation. Smad2 and Smad3 (the R-Smads) interact with and are phosphorylated by the type I receptor. Upon phosphorylation, the R-Smads form a complex with Smad4. The phospho-R-Smad/Smad4 complex translocates into the nucleus, binds to various proteins and DNA, and regulates transcription. The principal output of TGFβ signalling appears to be transcriptional regulation.

A distinctive feature of TGFβ signalling is the nuclear accumulation of the Smads. The regulation of Smad nuclear accumulation is poorly understood. Two mechanisms have been proposed: (*i*) the nucleocytoplasmic-shuttling-kinetics hypothesis and (*ii*) the retention-factor hypothesis [10]. The nucleocytoplasmic-shuttling-kinetics hypothesis states that the different forms of the Smads have different kinetics of nuclear import and export, such that the phosphorylated Smads accumulate in the nucleus. The retention-factor hypothesis states that binding factors in the cytoplasm have a higher affinity for unphosphorylated Smads whereas binding factors in the nucleus have a higher affinity for phosphorylated Smads. Candidate retention factors include the protein SARA (Smad anchor for receptor activation) [11, 12] and microtubules [13] in the cytoplasm and transcription factors and DNA in the nucleus [see reference 8 for a list of Smad-interacting transcription factors]. Since Smad nuclear accumulation is required for transcriptional regulation, our understanding of TGFβ signalling mechanisms depends on which hypothesis, if either, is correct.

While both hypotheses are reasonable, neither can completely account for existing data. For example, identical time courses of nuclear import for unmodified and phosphorylated Smads have been observed [14], which agrees with the more recent result that Smad3 and TGFβ-activated phospho-Smad3 are transported by a similar mechanism [15]. This work suggests that nucleocytoplasmic shuttling kinetics do not control Smad nuclear accumulation. Similarly, published data are not fully consistent with the idea that retention factors determine Smad signalling kinetics. First, the Smads cycle between the cytoplasmic and nuclear compartments at a significant rate, indicating that they are highly mobile within the cell. Treatment of cells with leptomycin-B, which blocks Smad4 nuclear export, induces constitutive nuclear localization of Smad4 within 10 min [16]. Second, strong tethering of phospho-R-Smads to nuclear retention factors during TGFβ signalling would limit the availability of the phospho-R-Smads for dephosphorylation. However, application of a TGFβ-type I receptor inhibitor during TGFβ signalling causes noticeable phospho-R-Smad dephosphorylation within minutes [17], indicating that the phospho-R-Smads are readily available for dephosphorylation. Third, existing evidence suggests that constitutive nucleocytoplasmic shuttling of the Smads allows for receptor activity monitoring [18], implying that a pool of Smads is freely mobile. Therefore, the retention-factor hypothesis cannot fully account for known Smad dynamics. The shortcomings of both hypotheses suggest that other unrecognised mechanisms might be important for determining Smad nuclear accumulation. Furthermore, various mechanisms could interact to effect Smad nuclear accumulation, making experimental determination difficult. The complexity of even the canonical TGFβ signalling pathway warrants a systems approach to this problem. Here we use mathematical modelling to understand the control of Smad nuclear accumulation.

Mathematical modelling has been used to analyse several cellular signalling systems including those that respond to EGF [19-21], PDGF [22], insulin [23], IFN-γ [24, 25], Wnt [26], and TNF-α [27]. Intracellular events in TGFβ signalling have not yet been modelled, although recent work has addressed signal processing at the receptor level [28]. Modelling serves to efficiently develop and/or assess competing hypotheses about complex signalling networks [e.g., above references, 29, 30, 31]. Modelling has also highlighted principles such as robustness [32, 33] and bistability [34, 35], and has been essential for the design of artificial signalling systems [36, 37]. While many cellular signalling systems have been modelled in isolation, the phenomenon of crosstalk will to lead to integrated models of multiple pathways. Such developments would be useful to the TGFβ field given that the effects of TGFβ are pluripotent and context-dependent, which may result from the multiple interactions of the Smads with other pathways [38-41]. Our model of TGFβ signalling is a useful first step in the development of an integrated multi-pathway model that includes the TGFβ pathway.

In this paper, we develop and analyse a dynamical model of Smad signalling. We sought to develop the simplest functional model that could reproduce the major features of Smad signalling. The model was based on data from the literature and our laboratory. We studied the effects of altering the model parameters on features of the model solutions. We conclude that the imbalance between R-Smad phosphorylation and dephosphorylation rates is likely an important determinant of Smad nuclear accumulation.

## 2. Materials and Methods

*2.1 Mathematical modelling*

Full justification for the model structure (Figure 1A), rate equations (Figure 1B), initial conditions (Table 1), and model input function (Supplementary Figure S1) is provided in the Supplementary Information. The model is a system of nonlinear ordinary differential equations based on conservation of mass [42]. The concentration of a species in the model is determined by the biochemical reactions in which it participates:

$$\frac{d[species]}{dt} = rate_{in} - rate_{out}.$$

For example, R-Smad-P$_{cyt}$ is produced by the phosphorylation of R-Smad$_{cyt}$ (reaction 1, Figure 1) and by the dissociation of the R-Smad-P/Smad4$_{cyt}$ complex (reaction 2, Figure 1) and is consumed by binding to Smad4$_{cyt}$ (reaction 2, Figure 1). The rates of reactions 1 and 2 are listed in Figure 1B. The mass balance for R-Smad-P$_{cyt}$ is therefore

$$\frac{d[R-Smad-P_{cyt}]}{dt} =$$

$$\frac{k_{cat} \cdot [receptors] \cdot [R-Smad_{cyt}]}{(K_1 + [R-Smad_{cyt}])} + k_{2d} \cdot [R-Smad-P/Smad4_{cyt}] - k_{2a} \cdot [R-Smad-P_{cyt}] \cdot [Smad4_{cyt}]$$

.

The system of ordinary differential equations was integrated numerically using the ode23s solver of Matlab (The MathWorks, Inc.). Statistical analysis was performed in Excel (Microsoft) and Matlab. The parameter sets and supporting raw calculations are provided as an Excel spreadsheet file in the Supplementary Information (referred to as the "Simulation supplement").

*2.1.1 Parameterisation.* None of the parameter values for the reactions included in the model are currently available in the literature. However, feasible ranges of the parameter values are known (Supplementary Table S1). These parameter ranges bounded distributions used in a computational sampling procedure inspired by the work of von Dassow et al. [33]. Specifically, we found randomly sampled parameter sets that gave qualitatively correct model solutions. The qualitative criteria were derived from the data of Pierreux et al. [16] (Section 2.1.2 includes a complete description of the criteria). After several rounds of testing, we obtained sufficient parameter sets for analysis (dataset1). Further details of the parameter sampling are provided in the Supplementary Information. Note that we constrained $k_{5cn}$ to equal $0.1 \times k_{5nc}$ in order to facilitate the recovery of the R-Smad$_{cyt}$ initial condition at the end of signalling and hence increase the probability of finding successful parameter sets.

*2.1.2 Dynamics of Smad signalling and criteria for model success.* Caroline Hill's group has published several time courses of Smad dynamics. In their most detailed time course, they measured the levels of phospho-Smad2 in the nuclei of HaCaT cells. The initially undetectable phospho-Smad2 increased to a maximum at ~45 min and then declined over 8 hours [18]. This pattern of phospho-R-Smad accumulation occurs for many cell types [e.g., 16, 18, 43-45], although the precise kinetics vary. For example, nuclear accumulation of phospho-R-Smads peaks after 15 min of TGFβ stimulation in untransformed mesangial cells [46].

Phosphorylation dynamics are correlated with the dynamics of downstream events. Both Smad complex formation during TGFβ and activin signalling [43, 47] and bulk Smad2 and Smad3 nuclear accumulation [18] show similar kinetics to those of Smad phosphorylation. Smad nuclear accumulation and exit is matched by Smad cytoplasmic exit and accumulation, respectively [16, 18]. Therefore Smad phosphorylation by the active receptor complex leads to downstream events such as Smad complex formation and nuclear accumulation. To constrain the model dynamics, we used the data of Pierreux et al. [16] to develop criteria for model success. Prior to listing the criteria, we define variables that correspond to those observed in the experimental data. Pierreux et al. [16] performed immunoblots of cytoplasmic and nuclear fractions using anti-Smad2 and anti-Smad4 antibodies. These antibodies recognize all forms of their target Smad molecules. Therefore, we defined the four variables: (1) $\Sigma\text{RSmad}_{cyt}$ (sum of R-$\text{Smad}_{cyt}$, R-Smad-$\text{P}_{cyt}$, and R-Smad-P/$\text{Smad4}_{cyt}$), (2) $\Sigma\text{RSmad}_{nucl}$ (sum of R-$\text{Smad}_{nucl}$, R-Smad-$\text{P}_{nucl}$, and R-Smad-P/$\text{Smad4}_{nucl}$), (3) $\Sigma\text{Smad4}_{cyt}$ (sum of $\text{Smad4}_{cyt}$ and R-Smad-P/$\text{Smad4}_{cyt}$), and (4) $\Sigma\text{Smad4}_{nucl}$ (sum of $\text{Smad4}_{nucl}$ and R-Smad-P/$\text{Smad4}_{nucl}$). The model outputs must satisfy five criteria: (1) The minimum of $\Sigma\text{RSmad}_{cyt}$ must occur between 15 and 180 min, with the value at the final time point (480 min) greater than the average of the maximum and minimum values. This forces the cytoplasmic R-Smad concentration to decrease to a minimum and then recover by the end of signalling. (2) The maximum of $\Sigma\text{RSmad}_{nucl}$ must occur between 15 and 180 min, with the value at the final time point less than 50% of the maximum value. (3) The minimum of $\Sigma\text{Smad4}_{cyt}$ must occur between 15 and 180 min, with the value at the final time point greater than the average of the maximum and minimum values. (4) The maximum of $\Sigma\text{Smad4}_{nucl}$ must occur between 15 and 180 min, with the level at the final time point less than 50% of the maximum value. (5) No negative or poorly scaled values are permitted: we require that the maximum number of molecules for any of the model state variables cannot be less than 10. We chose broad ranges of times for when the maxima or minima could occur because, as discussed above, these time ranges vary between cell types. In addition, this choice allowed us to capture a sufficiently large parameter ensemble.

*2.2 Experimental Procedures*

*2.2.1 Immunoblotting*. Cell lysates were prepared by growing 10 cm plates of cells to confluence in DMEM supplemented with 10% FBS, 100 IU/mL penicillin, 100 μg/mL streptomycin, and 2mM L-glutamine, followed by trypsinisation, counting, and lysis (Lysis buffer: 50 mM Tris-HCl, 400 mM NaCl, 1 mM EDTA, 1% NP-40, 15% glycerol, 2 mM sodium fluoride, 1 mM sodium orthovanadate, and protease inhibitors (Roche 1 697 498)). Protein concentrations were measured using the BCA assay (Pierce 23225). Lysates were separated by SDS-PAGE, transferred to a nitrocellulose membrane, and immunoblotted using antibodies against Smad2 (Zymed 51-1300), phospho-Smad2 (a gift from Dr. Aris Moustakas), Smad3 (Zymed 51-1500), Smad4 (B-8, Santa Cruz sc-7966), or α-tubulin (MP Biomedicals 691251). Detection was performed using either chemifluorescence (ECL Plus, Amersham RPN 2132) or chemiluminescence (Supersignal West Dura Extended Duration Substrate, Pierce 34076). Exposures were performed for different lengths of time to ensure that band intensities fell into the linear range.

*2.2.2 Quantitative immunoblotting*. Quantitative immunoblots were performed by comparing the band intensities of cell lysates to a standard curve of recombinant protein dilutions. Recombinant GST-fused human Smad2, Smad3, and Smad4 were purified from *E. coli* using standard procedures, the expression vectors for which are described in Zhang et al. [48]. The concentrations of the purified proteins were determined by separating the proteins alongside a series of BSA dilutions of known concentration (Pierce), followed by Coomassie staining and densitometry. The recombinant proteins were serially diluted and separated alongside the cell lysates by SDS-PAGE followed by immunoblotting as described above. The diluted recombinant protein was use to generate a standard curve and the cellular protein content was interpolated from those curves. Raw data and calculations for the quantitative immunoblotting experiment are provided in spreadsheet form as Supplementary Figure S2, with the live spreadsheet posted online ("Smad quantification supplement").

*2.2.3 TGFβ time course*. Cells were grown to confluence in 10 cm dishes. Five millilitres of media supplemented with 50 pM TGFβ was applied to the cells for the indicated times. At the end of the time period, cells were rinsed twice with phosphate-buffered saline and frozen with liquid nitrogen. Cells were stored at -80$^o$C until lysis.

## 3. Results

*3.1 Smad quantification assays*. We estimated the number of cellular Smad proteins by quantitative immunoblotting of Mv1Lu cell lysates. Confidence intervals for the cellular molecule numbers are provided in Table 2 with raw data and calculations supplied as Supplementary Information (Figure S2, Smad quantification supplement). We observed that Smad2 and Smad4 were present in similar amounts, while Smad3 was present in amounts 4- to 5-times less than Smads 2 and 4. This finding was consistent across several cell lines (Supplementary Figure S3, Table S2). These results were used to establish the model initial conditions (Table 1). To account for the larger abundance of Smads in several TGFβ-sensitive cell lines (Supplementary Figure S3, Table S2, reference [49]), the initial conditions were set to higher total numbers of Smads than our estimates. We note that the model behaviour is robust to changes in the initial conditions (data not shown).

*3.2 The model qualitatively reproduces Smad signalling dynamics*. We determined sets of parameter values that led to qualitatively correct model outputs. We randomly sampled parameters sets and found 203 sets that led to solutions satisfying all 5 criteria (successful parameter sets are listed in the Simulation supplement). Supplementary Figure S4 shows the 203 curves associated with the first four criteria, demonstrating that the model faithfully reproduces the qualitative trends from the published data [16] over a range of biologically plausible parameter values. We show a representative result in Figure 2A in which the model was run using the median parameter values (Table 3). To provide further evidence that our model outputs were qualitatively correct, we performed a time course of Smad2 phosphorylation in mink lung epithelial cells in response to TGFβ stimulation. Figure 2B shows a representative plot of ΣR-Smad-P (the sum of all phospho-R-Smad species in the cell) versus time with the experimental data shown below. The model captures the trends in the data. Also, the Smad4 blot validates our assumption of constant Smad4 concentration during signalling.

*3.3 Parameter sensitivity and correlations*. We used our set of successful parameters to study the sensitivity of the model behaviour to the parameter values. We first examined the ranges of each parameter in dataset1. If a parameter must fall within a relatively narrow range for the model to be successful, the model is more sensitive to changes in the value of that parameter. As shown in Table 3 (rightmost column), most of the parameter values varied by more than 4 orders of magnitude. The two exceptions were $k_{cat1}$, the turnover number of the receptor kinase, which was distributed over only 2.84 orders of magnitude, and $k_{6d}$, the dissociation rate constant for the nuclear Smad complex, which was distributed over 3.04 orders of magnitude. The qualitative performance of the model is therefore most sensitive to $k_{cat1}$ and $k_{6d}$.

Proper model behaviour may depend on ratios of parameters. To examine this possibility, we calculated pairwise correlations for all the parameters. Two pairs of parameters were highly significantly correlated ($P<10^{-16}$): (1) $k_{cat1}$ and $K_1$ and (2) $k_{6a}$ and $k_{6d}$ (the full correlation matrix is provided in Supplementary Table S3). The median ratio for $k_{cat1}/K_1$ was $1.30\times10^{-5}$ cell·molecule$^{-1}$·min$^{-1}$ and for $k_{6d}/k_{6a}$ was 508 molecules·cell$^{-1}$. Appropriate model behaviour requires that the parameters determining the rates of Smad phosphorylation and nuclear reversible complex formation be constrained to those ratios.

The receptor catalytic efficiency, $k_{cat1}/K_1$, is within range of known values for enzymes [50] (see the Simulation supplement for calculations). The ratio $k_{6d}/k_{6a}$ is equivalent to the dissociation constant ($K_d$) of phospho-R-Smad and Smad4. The published estimate for the $K_d$ of active Smad2 and Smad4 is 79 nM [51]. Assuming a nuclear volume of $10^{-13}$ L [52], conversion of 508 molecules·cell$^{-1}$ to units of mol·L$^{-1}$ gives ~8 nM, which compares reasonably well with the published estimate. Similarly, the median ratio for $k_{2d}/k_{2a}$, which corresponds to the same reaction as $k_{6d}/k_{6a}$ except in the cytoplasm (reaction 2), was ~1 nM. We emphasize that no *a priori* constraints were placed on any of $k_{2a}$, $k_{2d}$, $k_{6a}$, and $k_{6d}$ during the parameter search procedure. This result implies that the parameter search procedure obtained realistic parameter values.

Together, the descriptive statistics suggest that the qualitative behaviour of the model was affected mainly by $k_{cat1}$ and $k_{6d}$.

*3.4 Regression analysis of parameter effects.* To quantitatively predict how each parameter contributes to Smad nuclear accumulation, we employed multiple regression analysis. We generated a second dataset, dataset2, which consisted of the predictor and response variables. The predictor variables were the model parameters. Parameter values were sampled orthogonally by selecting 2 values per parameter, the 25[th] and 75[th] quartile values (Table 3), and using those values to generate parameter sets containing every combination of the two values for each parameter. Because $k_{5cn}$ was constrained to equal $0.1\times k_{5nc}$, it was not included in the analysis. Dataset2 therefore included $2^{12} = 4096$ parameter sets (Simulation supplement). An orthogonal design is one in which the dot product of any two columns in the design matrix sum to 0 (here the factor levels would be assigned -1 or +1). Orthogonal designs imply that the effects of each term in the regression model are decoupled, thus allowing for unbiased analysis in which the estimators of the regression coefficients are uncorrelated [53]. Two response variables were analysed: the time integrals of $\sum RSmad_{nucl}$ (denoted $\int\sum RS_{nucl}$) and $\sum Smad4_{nucl}$ ($\int\sum S4_{nucl}$). The integral is the area under the time-course curve and corresponds to the total number of Smads that pass through the nucleus during signalling. This quantity is considered a key determinant of TGFβ signalling output [10, 54], so our response variables are biologically relevant. To determine the response variables, we ran the model with the new parameter sets and calculated the time integrals numerically using the trapezoidal rule. Most model outputs were qualitatively consistent with experimental time course data, but not all met our criteria for model success (see Supplementary Figure S5 for the time curves associated with dataset2).

The regression model terms included all main effects (each of the parameters alone) and pairwise (the products of all possible parameters pairs) and three-way (the products of all possible combinations of three parameters) interaction effects. Higher-order interaction effects were used as the error estimate [53]. Complete results and residual diagnostic plots for the full regression model are included in the Simulation supplement. We observed that the main effect and binary and ternary interaction terms account for almost all of the response variable variance: the adjusted $R^2$ values for the full regression models were 98.48% for $\int\Sigma RS_{nucl}$ and 98.46% for $\int\Sigma S4_{nucl}$ (Table 4). This result implies that our orthogonal sampling strategy was effective.

To determine the model parameter sensitivities, we compared the predictive ability of different regression models (quantified by the adjusted $R^2$ value) in which each of the predictor variables was eliminated one-at-a-time from the full regression model. The more important a predictor variable is, the more the adjusted $R^2$ decreases upon removal of the variable from the full regression model. The removal of either of the two parameters associated with R-Smad phosphorylation, $k_{cat1}$ and $K_1$, caused the greatest decrease in the adjusted $R^2$ for both $\int\Sigma RS_{nucl}$ and $\int\Sigma S4_{nucl}$ (Table 4). Removal of $k_{6d}$ also had substantial impact on the adjusted $R^2$ values for both $\int\Sigma RS_{nucl}$ and $\int\Sigma S4_{nucl}$. The parameters $v_{max7}$ and $K_7$ were important determinants of the adjusted $R^2$ value for $\int\Sigma RS_{nucl}$, while $k_{4nc}$ and $k_{6a}$ were important determinants of the adjusted $R^2$ value for $\int\Sigma S4_{nucl}$. The removal of $k_{2a}$, $k_{2d}$, $k_3$, $k_{4cn}$, or $k_{5nc}$ had either no effect or even a positive effect on the adjusted $R^2$ for both $\int\Sigma RS_{nucl}$ and $\int\Sigma S4_{nucl}$. Together, these results suggest that the rate of R-Smad phosphorylation is the most important determinant of overall Smad nuclear accumulation. To a lesser extent, parameters involved in reversible Smad complex formation in the nucleus ($k_{6a}$, $k_{6d}$), R-Smad dephosphorylation ($v_{max7}$, $K_7$), and Smad4 availability in the nucleus ($k_{4nc}$) are also important determinants of Smad nuclear accumulation. To further explore the roles of these parameters in Smad dynamics, we examined the reaction rates.

*3.5 Analysis of rates.* The rate of change in concentration of a species is determined by the rates of the reactions in which it is involved. We calculated the reaction rates by substituting the concentrations of the species at each time point into the rate equations (Figure 1B). Figure 3A shows the absolute values of the reaction rates for the median parameter set. We note that only 2 curves are labelled, because all the rates tend to approach either curve 1 (reactions 1, 2, and 3) or curve 2 (reaction 4, 5, 6, and 7). Reactions 1, 2, and 3 represent R-Smad phosphorylation, reversible Smad complex formation in the cytoplasm, and Smad complex nuclear import, respectively. Reactions 4, 5, 6, and 7 represent Smad4 nuclear shuttling, R-Smad nuclear shuttling, reversible Smad complex formation in the nucleus, and phospho-R-Smad dephosphorylation. Curves 1 and 2 intersect at the time roughly corresponding to maximal Smad nuclear accumulation (Figure 3B), with the magnitudes of the rates associated with curve 1 initially higher than those of curve 2. This suggests that the time at which the rates intersect – that is, the time when the rate of R-Smad dephosphorylation first becomes larger than the rate of R-Smad phosphorylation – may predict the time of maximal Smad nuclear accumulation. To test this possibility, we calculated the time at which the rate of dephosphorylation (reaction 7) was equal to that of R-Smad phosphorylation (reaction 1) and the time of maximal nuclear $\Sigma$R-Smad concentration using the parameters in dataset1. Similarly, we calculated the time at which the rate of Smad complex formation (reaction 6) intersected that of R-Smad phosphorylation and the time of maximal nuclear ΣSmad4 concentration (raw numbers are provided in the Simulation supplement). In both cases, the times were correlated (Figures 3C and 3D).

It appears that the rate of phosphorylation sets the overall rate of cytoplasmic events. Initially, this rate is higher than the rates of nuclear reactions, which promotes Smad nuclear accumulation. The decrease of receptor activity reduces the rates of the cytoplasmic reactions while the increase in Smad concentration in the nucleus increases the rates of the nuclear reactions. This continues until the rates of the nuclear reactions overtake those of the cytoplasm. The Smads then relocalise to the cytoplasm. However, we have not yet addressed which of the nuclear reactions is principally responsible for setting the rates of the other nuclear reactions.

*3.6 Dephosphorylation could be the rate-limiting reaction in the nucleus.* Our results suggest that the reactions principally controlling Smad nuclear accumulation are (*i*) the phosphorylation of the cytoplasmic R-Smads by active receptors, (*ii*) reversible Smad complex formation in the nucleus and, (*iii*) the dephosphorylation of phospho-R-Smads in the nucleus. To further evaluate the roles of reversible Smad complex formation and R-Smad dephosphorylation as controlling steps in Smad signalling, we studied the model without Smad4. Smad4 is often inactivated in colon and pancreatic cancer cell lines [4]. However, the kinetics of R-Smad phosphorylation and nuclear translocation are not necessarily affected by the loss of Smad4 [43, 55]. To see whether the model could replicate R-Smad signalling in Smad4-null cell lines, we adjusted the initial condition of Smad4. For all results shown in Figures 4 and 5, the model was altered to include reaction 8, which describes the nuclear import reaction for R-Smad-$P_{cyt}$ (Figure 1). We used the median parameter set (Table 3) and set the rate constant of phospho-R-Smad nuclear import ($k_8$) equal to the median rate constant of nuclear import for the phospho-R-Smad/Smad4 complex ($k_3$=16.6 min$^{-1}$). Figure 4A shows the effect of reducing the cellular Smad4 concentration on the simulated time course of ΣR-Smad nuclear accumulation. In contrast to published data [43], R-Smad nuclear accumulation is lost with decreasing Smad4. To determine if different parameter values for R-Smad phosphorylation and dephosphorylation could force the model to replicate the known behaviour of the R-Smads with Smad4 deletion, we tested all combinations of 4 different values for each of $v_{max7}$, $K_7$, $k_{cat1}$, and $K_1$ (values are listed in Supplementary Table S4 and in the Simulation supplement), with the other parameter values set at their median values. The resulting 256 curves show that the model is capable of replicating R-Smad kinetics with Smad4 deleted (Figure 4B). The curves highlighted in blue are those that satisfy the first, second, and fifth criteria for model success (76/256 curves). We calculated the means of the 76 values for each of the four adjusted parameters and used those values as the basis for a new parameter set, in which the remaining parameters were set to their median values from dataset1 (Table 5). Running the model with the new parameter set revealed less decrease of nuclear R-Smad accumulation under conditions of decreasing Smad4 (Figure 4C). Therefore, slow dephosphorylation can preserve R-Smad nuclear accumulation independent of Smad4 concentration. We conclude that phospho-R-Smad dephosphorylation in the nucleus could be a viable mechanism for Smad nuclear accumulation.

*3.7 Is slow phospho-R-Smad dephosphorylation required for Smad nuclear accumulation?* Rate-limiting phospho-R-Smad dephosphorylation is sufficient for Smad nuclear accumulation in our model; we asked whether rate-limiting dephosphorylation is also necessary for Smad nuclear accumulation. To estimate the contribution of phospho-R-Smad dephosphorylation to Smad nuclear accumulation, we examined the limiting behaviour of very rapid phospho-R-Smad dephosphorylation. This allowed us to test whether R-Smad nuclear accumulation can occur solely by reversible complex formation with binding factors, represented in our model by Smad4.

The parameter sets and Smad4 initial conditions associated with Figure 4D are listed in Table 5. The baseline parameter set includes the new values for the phosphorylation and dephosphorylation parameters found in the analysis of Figure 4B. We also adjusted the values of $k_{2a}$, $k_{2d}$, $k_{6a}$, and $k_{6d}$ so that the phospho-R-Smad/Smad4 complex $K_d = 79$ nM [51] in both the cytoplasm and nucleus. In addition, we adjusted $k_{4cn} = 0.2 \times k_{4nc}$ to reflect the Smad4 distribution under basal conditions. Figure 4D, curve 1 shows the simulated time course of $\sum$R-Smad$_{nucl}$ using the baseline parameter set.

Our first test examined R-Smad nuclear accumulation under conditions of constant rapid nuclear phospho-R-Smad dephosphorylation. We set $v_{max7} = 10^6$ molecules·cell$^{-1}$·min$^{-1}$ and $K_7 = 1$ molecule·cell$^{-1}$, with the other parameters unchanged from baseline (Table 5). This change eliminates R-Smad nuclear accumulation (Figure 4D, curve 2). Therefore, Smad4 cannot solely mediate Smad nuclear accumulation when dephosphorylation is rapid because its affinity for the R-Smads is too low. Next, we determined whether adjusting the affinity of the Smad complex could restore R-Smad nuclear accumulation. Changing $k_{2a}/k_{2d}$ and $k_{6a}/k_{6d}$ to give an equilibrium constant of 1 pM, which is well below the estimated $K_d$ of 79 nM [51], led to substantial R-Smad nuclear accumulation but still less than with slow dephosphorylation (Figure 4D, curve 3). Therefore, reversible interactions of the Smads with each other could in principle account for Smad nuclear accumulation. However, the unphysically high affinity required between the Smads confirms that Smad4 is unlikely to act as the sole factor that protects the R-Smads from the nuclear phosphatase.

Constitutive nuclear shuttling causes a significant proportion of Smad4 to localise to the cytoplasm, even during signalling. We therefore asked whether confining Smad4 to the nucleus would enhance its binding to the phospho-R-Smads and increase Smad nuclear accumulation. To carry out this test, we sequestered all of the Smad4 molecules in the nucleus by setting $k_{4cn} = 10^4$ min$^{-1}$ and $k_{4nc} = 0$, and set the initial condition of Smad4$_{cyt} = 0$ and that of Smad4$_{nucl} = 1.5 \times 10^5$ molecules (Table 5). The remaining parameters were the same as those used to produce curve 3, including the Smad complex $K_d$ of 1 pM. To our surprise, we found that under conditions of rapid dephosphorylation, R-Smad nuclear accumulation decreased when Smad4 was localised to the nucleus (Figure 4, curve 4). This result is more dramatic when the affinity of the binding factor is further reduced. We set $k_{2a}/k_{2d}$ and $k_{6a}/k_{6d}$ to 1 nM, which still represents a higher affinity than that experimentally observed for active R-Smad and Smad4. We ran the model with Smad4 normally localised and with Smad4 confined to the nucleus. If the binding factor is localized to the cytoplasm, then R-Smad nuclear accumulation occurs, although less so than when the affinity is 1 pM (Figure 4D, curve 5). If the binding factor is localised to the nucleus, no R-Smad nuclear accumulation occurs (Figure 4D, curve 6).

Given the reduced ability of constitutively nuclear Smad4 to promote Smad nuclear accumulation, we examined what the concentration requirement would be for retention factors that bind the R-Smads with lower affinity. One possible retention factor is DNA, which is thought to be a candidate for Smad tethering [10, 56]. The affinity of the Smad3 MH1 domain, which contains the Smad DNA binding motif, for a sequence of DNA containing the putative Smad binding element (SBE, 5'-GTCT-3') is approximately 100 nM [57]. We used nuclear-localised Smad4 to represent the DNA binding sites and set the $K_d$ of phospho-R-Smad and Smad4 to 100 nM (Table 5). We then solved the model with increasing amounts of Smad4 and plotted the corresponding $\Sigma$R-Smad$_{nucl}$. When dephosphorylation is rapid, detectable amounts of R-Smad nuclear accumulation above the basal nuclear amount (Figure 5, curve 1) required $\sim 15 \times 10^8$ binding sites (Figure 5, curve 2). Achieving maximal nuclear accumulation corresponding to slightly less than 50% of the total R-Smads ($>8 \times 10^4$) in the nucleus required $\sim 63 \times 10^9$ nuclear binding sites (Figure 5, curve 4). Since only $6 \times 10^9$ base pairs constitute the entire human genome [58, p. 222], DNA is unlikely to be the sole mediator of Smad nuclear retention. Finally, we repeated the analysis except we let the nuclear factors bind phospho-R-Smad with $K_d = 1$ pM. Even with such tight binding, $\sim 7.5 \times 10^5$ binding sites would be required to achieve $\sim 8 \times 10^4$ R-Smads maximally accumulated in the nucleus during signalling (Supplementary Figure S6). Therefore, if a retention factor is responsible for Smad nuclear accumulation when dephosphorylation is rapid, the factor must have both high affinity and high abundance. We conclude that rate-limiting phospho-R-Smad dephosphorylation is at least partially responsible for Smad nuclear accumulation.

## 4. Discussion

An important feature of TGF$\beta$ signalling is the nuclear accumulation of the Smad proteins, which leads to transcriptional regulation of TGF$\beta$-target genes. However, the mechanism by which the Smads accumulate in the nucleus is not well understood. Two hypotheses exist: the nucleocytoplasmic-shuttling-kinetics hypothesis and the retention-factor hypothesis [10]. However, neither hypothesis can fully account for published data. We used mathematical modelling to examine the existing hypotheses and to explore alternatives. We conclude that the imbalance in the rates of R-Smad phosphorylation in the cytoplasm and dephosphorylation in the nucleus likely significantly contributes to Smad nuclear accumulation.

An initial obstacle to our modelling approach was that the kinetic rate constants have not previously been measured. We addressed this problem by sampling parameters over a broad range of biologically plausible values. Rather than determining one best-fit parameter set, we analysed the statistics of many parameter sets that produce qualitatively correct model behaviour. Related work studying ensembles of parameters has been done in models of other biochemical signalling networks [24, 33, 59-63]. Further development and refinement of these methods is necessary as biological models expand in scope.

Our parameter sensitivity analysis revealed that changes in the parameters involved in R-Smad phosphorylation and dephosphorylation and in reversible Smad complex formation in the nucleus most affected the degree of Smad nuclear accumulation. In contrast, we observed in the regression analysis that $k_3$, $k_{4cn}$, and $k_{5nc}$, all nuclear shuttling parameters, were relatively unimportant in determining Smad nuclear accumulation. Our results are corroborated by recent data showing that different nuclear import/export rates cannot explain TGF$\beta$-mediated Smad nuclear accumulation [64]. Hence, the nucleocytoplasmic-shuttling-kinetics hypothesis likely does not account for Smad nuclear accumulation.

Further insight into the dynamics of Smad signalling resulted from the analysis of reaction rates. We found that the rates of cytoplasmic reactions tend to approach the same rate, as do the rates of nuclear reactions. During signalling, the rates of the cytoplasmic reactions are initially high but decrease as receptor activity decreases, eventually becoming less than the rates of the nuclear reactions. The time when the cytoplasmic and nuclear reaction rates are equal predicts the time of maximal Smad nuclear accumulation. Because R-Smad phosphorylation is the first intracellular step in signalling and because the rate constants associated with this reaction, $k_{cat1}$ and $K_1$, are sensitive parameters, we propose that the rate of R-Smad phosphorylation sets the overall rate of signalling.

We then sought to identify which nuclear reaction was rate-limiting. Since the parameters involved in two nuclear reactions (Smad complex formation and R-Smad dephosphorylation) were identified as important, either a slow phosphatase or Smad4-mediated sequestration of the R-Smads from dephosphorylation could control Smad nuclear accumulation. To distinguish these possibilities, we studied the effects of changing the parameter values and Smad4 abundance and/or localisation. Either a slow phosphatase or Smad4-mediated sequestration alone could in principle mediate Smad nuclear accumulation. Specifically, when Smad4 is deleted, slow dephosphorylation can lead to Smad nuclear accumulation. Similarly, if dephosphorylation is rapid, then changing the binding affinity of the phospho-R-Smad/Smad4 complex can also lead to R-Smad nuclear accumulation. In the latter case, however, we observed that the required $K_d$ for the Smad complex was ~1 pM, which is considerably less than the reported estimate of 79 nM [51]. This suggests that the affinity between phospho-R-Smad and Smad4 is not sufficiently high to permit Smad4 to protect the phospho-R-Smads from rapid dephosphorylation in the nucleus. This result supports the notion that dephosphorylation is at least partially responsible for Smad nuclear accumulation.

To confirm that dephosphorylation must be at least partially rate-limiting, we tested the effects of binding factors localised exclusively in the nucleus. (When the binding factors are nuclearly localised, we alter the model to allow Smad4-independent nuclear import of phospho-R-Smad). We expected that restricting the binding factors to the nucleus would enhance Smad nuclear accumulation and therefore make dephosphorylation less important. Instead, nuclear-localised binding factors decrease R-Smad nuclear accumulation when dephosphorylation is rapid. In this case, the phosphatase rapidly dephosphorylates the imported unbound phospho-R-Smad before it can interact with a binding factor. This result implies that if nuclear retention factors are the principal mechanism for Smad nuclear accumulation, then such factors should either be present at high abundance or they should bind the phospho-R-Smads with high affinity. Therefore nuclear retention factors are unlikely to be the sole mechanism of Smad nuclear accumulation. Instead, we propose that an imbalance in the rates of R-Smad phosphorylation and dephosphorylation must significantly contribute to Smad nuclear accumulation.

At present, our model cannot address one alternative scenario that could support the retention-factor hypothesis: oligomerisation of the Smads into multimeric complexes could be important. We made the simplifying assumption that the phospho-R-Smads form heterodimeric complexes with Smad4. Our assumption is justified from experimental evidence showing that R-Smads and Smad4 form heterodimers [51, 65, 66]. However, we ignore the possible effects of phospho-R-Smad homodimers and phospho-R-Smad/Smad4 heterotrimers, which have been shown to occur in structural studies *in vitro* and upon overexpression *in vivo* [67, 68]. In future theoretical work we plan to address whether participation of the Smads in larger complexes can better sequester the phospho-R-Smads from dephosphorylation. This issue is particularly relevant given our finding that complexes formed in the cytoplasm more effectively sequester the phospho-R-Smads from rapid dephosphorylation than factors localised exclusively in the nucleus. In this way, the Smads could act as their own nuclear retention factors.

In addition, a similar phenomenon could occur in the nucleus given the number of transcription factors known to interact with the Smads [8]. Formation of multiple types of complexes in the nucleus could also significantly modulate the kinetics of Smad nuclear accumulation. Indeed, indirect evidence of these complexes occurring *in vivo* exists as the mobility of overexpressed fluorescent-protein-tagged Smad2 in the nucleus is substantially decreased upon TGFβ signalling [56, 64]. The reduction in mobility was interpreted as regulated tethering of the Smads in the nucleus, suggesting that nuclear retention is the mechanism of Smad nuclear accumulation [64]. However, other work suggests that transcriptional activation complexes are highly dynamic: components are continually binding to and dissociating from the complex [69]. If participation of individual Smad molecules in large complexes is transient, then such complexes are unlikely to sequester the phospho-R-Smads from rapid dephosphorylation. Moreover, participation of the Smads in complexes is not mutually exclusive with the hypothesis that the principal mechanism of Smad nuclear accumulation is an imbalance in R-Smad phosphorylation and dephosphorylation rates. Smad nuclear accumulation mediated by slow dephosphorylation could drive reversible interactions involving the Smads in the direction of complex formation, thus promoting the ability of the Smads to regulate transcription. At this time, little data is available on the copy numbers and affinities for the Smads of transcription factors and other putative nuclear binding partners. Therefore, we are currently unable to specify how much retention factors versus an imbalance in phosphorylation/dephosphorylation rates contribute to Smad nuclear accumulation. We do conclude, however, that both mechanisms are likely to significantly contribute to Smad nuclear accumulation.

We integrate our observations to propose the following sequence for TGFβ signalling: Ligand binding rapidly activates the receptor kinases and the rate of Smad phosphorylation quickly increases and surpasses the rate of dephosphorylation. The Smads form complexes in the cytoplasm that are imported into the nucleus. The Smad complexes accumulate in the nucleus because the phosphatase cannot dephosphorylate the phospho-R-Smads rapidly enough to prevent accumulation. This is due to an intrinsically slow phosphatase but also to sequestration of the phospho-R-Smad from the phosphatase by participation in complexes. However, as receptors are downregulated, the rate of phosphorylation eventually decreases below that of dephosphorylation. At this time, the accumulation of Smads in the nucleus begins to subside and returns to basal levels.

Our model leads to important suggestions for the control of Smad dynamics and for the design of therapies for diseases involving TGFβ signalling. The cell has various means to regulate Smad signalling: the expression levels of the receptors, the nuclear phosphatase, and binding partners in the cytoplasm and nucleus could be balanced to ensure proper Smad signalling kinetics. (Note that not all cells necessarily adjust these expression levels to maintain Smad dynamics, as Smad nuclear accumulation in certain cancer cell lines expressing low levels of TGFβ receptors is of shorter duration [43]). Nonetheless, artificial targeting of phosphatase expression and/or activity could represent a novel potential target for therapy in diseases involving Smad signalling. Our model predictions therefore stress the need to identify the nuclear phosphatase. In addition, our results suggest that if dephosphorylation is rapid, then expressing a binding factor that forms a complex with phospho-R-Smad in the cytoplasm may serve as a better nuclear retention factor than the same factor constitutively localised in the nucleus.

In summary, we have constructed and analysed a mathematical model of Smad signalling. Our findings lead us to conclude that the imbalance between R-Smad phosphorylation and dephosphorylation rates is likely a significant mechanism of Smad nuclear accumulation during TGFβ signalling. We are currently investigating this hypothesis experimentally.

## 5. Acknowledgements

We thank Dr. David Clough, Maribeth Oscamou, the editor, Jacky Snoep, and the two anonymous reviewers for helpful comments. D. C. C. was supported by a postgraduate scholarship from the Natural Sciences and Engineering Research Council of Canada and a Doctoral Research Award from the Canadian Institutes of Health Research. M. D. B. was supported by a Research Fellowship from the Alfred P. Sloan Foundation. This work was supported by an NIH grant to X. L. (CA107098-01).

## Figure Legends

**Figure 1**. (A) Schematic of our Smad signalling model. Reactions are numbered arbitrarily. Reaction 8 was included only for model variants used in Figures 4 and 5 to account for Smad4-independent phospho-R-Smad nuclear import. (B) Model equations. Equations for cytoplasmic and nuclear phospho-R-Smad were modified as shown (lower right) when reaction 8 was included in the model.

**Figure 2**. Representative model solution. (A) Model solution generated using the median parameter values. The curves represent the number of molecules per cell of (1) $\sum$R-Smad$_{nucl}$, (2) $\sum$Smad4$_{nucl}$, (3) $\sum$R-Smad$_{cyt}$, and (4) $\sum$Smad4$_{cyt}$. (B) Model simulations compare well with experimental data. The plot shows the simulated time course for $\sum$R-Smad-P (cellular phospho-R-Smad) generated using the median parameter set. The immunoblot below shows the relative levels of phospho-R-Smad in mink lung epithelial cells stimulated with 50 pM TGFβ. The trends in the data compare well with the simulation. In the experiment, the membrane was probed for Smad4, to verify that the levels of Smad4 are constant for the duration of signalling, and for α-tubulin, a loading control.

**Figure 3**. Reaction rates. (A) Reaction rates generated with the median parameter set. The curves show the rates of (1) reactions 1, 2, and 3 and (2) reactions 4, 5, 6, and 7. (B) The time of intersection of (1) phosphorylation and (2) dephosphorylation rates is correlated with the time of maximal accumulation of the nuclear R-Smads (3), for the solution generated with the median parameter set. (C) and (D) The intersection time of reaction rates 1 and 7 correlates with the time of maximal R-Smad nuclear accumulation (C) while the intersection time of reaction rates 1 and 6 correlates with the time of maximal Smad4 nuclear accumulation (D). The model parameters from dataset 1 were used to calculate the times at which the rates intersected and the times of maximal Smad nuclear accumulation.

**Figure 4**. Slow dephosphorylation can mediate Smad nuclear accumulation. (A) The effect of reductions in Smad4 concentration on R-Smad nuclear accumulation. Curves of $\sum$R-Smad were generated using the median parameter set (Table 3) and total Smad4 cellular concentrations (molecules·cell$^{-1}$) of: (1) $15 \times 10^4$, (2) $10 \times 10^4$, (3) $5 \times 10^4$, and (4) 0. Note that reaction 8, which describes the nuclear import of phospho-R-Smad, was included in the model used to generate all of the curves in Figure 4. The corresponding parameter, $k_8$, was set to the median value of $k_3$. (B) Effect of varying phosphorylation and dephosphorylation parameters on R-Smad nuclear accumulation under conditions of Smad4 deletion. All combinations of 4 levels of $k_{cat1}$, $K_1$, $v_{max7}$, and $K_7$ (Table S4) were studied; the remaining parameters were set to their median values and the concentration of Smad4 was set to 0. Outputs satisfying criteria 1, 2, and 5 (the R-Smad criteria) are coloured blue (76 out of 256 curves). (C) The effect of reducing Smad4 concentration on R-Smad nuclear accumulation using adjusted parameters for R-Smad phosphorylation and dephosphorylation (Table 5). The model was integrated with four levels of total cellular Smad4 concentration (molecules·cell$^{-1}$): (1) $15 \times 10^4$, (2) $10 \times 10^4$, (3) $5 \times 10^4$, and (4) 0. (D) The effect of the dephosphorylation rate, phospho-R-Smad and Smad4 binding affinity, and R-Smad binding factor localisation on R-Smad nuclear accumulation. The different curves show $\sum$R-Smad$_{nucl}$ as a function of time for different variants of the model (for parameter values, see Table 5): (1) baseline parameters; (2) rapid dephosphorylation; (3) $K_d$ of phospho-R-Smad and Smad4 set to 1 pM to compensate for the rapid dephosphorylation; (4) $K_d$ of phospho-R-Smad and Smad4 set to 1 pM and constitutive nuclear localisation of Smad4; (5) $K_d$ of phospho-R-Smad and Smad4 set to 1 nM; (6) $K_d$ of phospho-R-Smad and Smad4 set to 1 nM and constitutive nuclear localisation of Smad4.

**Figure 5**. Effect of nuclear binding factor concentration on R-Smad nuclear accumulation. The curves correspond to time course simulations of $\sum$R-Smad$_{nucl}$ in which nuclear binding factor concentrations were set to (1) $15 \times 10^6$, (2) $15 \times 10^8$, (3) $25 \times 10^9$, and (4) $63 \times 10^9$ molecules·cell$^{-1}$. The K$_d$ of phospho-R-Smad and Smad4 was set to 100 nM (for parameter values, see Table 5).

**Figure 1A**.

$$\frac{d[receptors]}{dt} = -100 \cdot e^{\left(-t/90\right)}$$

$$\frac{d[R-Smad_{cyt}]}{dt} = r5 - r1$$

$$\frac{d[R-Smad-P_{cyt}]}{dt} = r1 - r2$$

$$\frac{d[Smad4_{cyt}]}{dt} = r4 - r2$$

$$\frac{d[R-Smad-P \cdot Smad4_{cyt}]}{dt} = r2 - r3$$

$$\frac{d[R-Smad-P \cdot Smad4_{nucl}]}{dt} = r3 - r6$$

$$\frac{d[R-Smad_{nucl}]}{dt} = r7 - r5$$

$$\frac{d[R-Smad-P_{nucl}]}{dt} = r6 - r7$$

$$\frac{d[Smad4_{nucl}]}{dt} = r6 - r4$$

$$r1 = \frac{k_{cat1} \cdot [receptors] \cdot [R-Smad_{cyt}]}{\left(K_1 + [R-Smad_{cyt}]\right)}$$

$$r2 = k_{2a} \cdot [R-Smad-P_{cyt}] \cdot [Smad4_{cyt}] - k_{2d} \cdot [R-Smad-P \cdot Smad4_{cyt}]$$

$$r3 = k_3 \cdot [R-Smad-P \cdot Smad4_{cyt}]$$

$$r4 = k_{4nc} \cdot [Smad4_{nucl}] - k_{4cn} \cdot [Smad4_{cyt}]$$

$$r5 = k_{5nc} \cdot [R-Smad_{nucl}] - k_{5cn} \cdot [R-Smad_{cyt}]$$

$$r6 = k_{6a} \cdot [R-Smad-P_{nucl}] \cdot [Smad4_{nucl}] - k_{6d} \cdot [R-Smad-P \cdot Smad4_{nucl}]$$

$$r7 = \frac{v_{max7} \cdot [R-Smad-P_{nucl}]}{\left(K_7 + [R-Smad-P_{nucl}]\right)}$$

$$r8 = k_8 \cdot [R-Smad-P_{cyt}]$$

Modified equations when reaction 8 included in model:

$$\frac{d[R-Smad-P_{cyt}]}{dt} = r1 - r2 - r8$$

$$\frac{d[R-Smad-P_{nucl}]}{dt} = r6 - r7 + r8$$

**Figure 1B.**

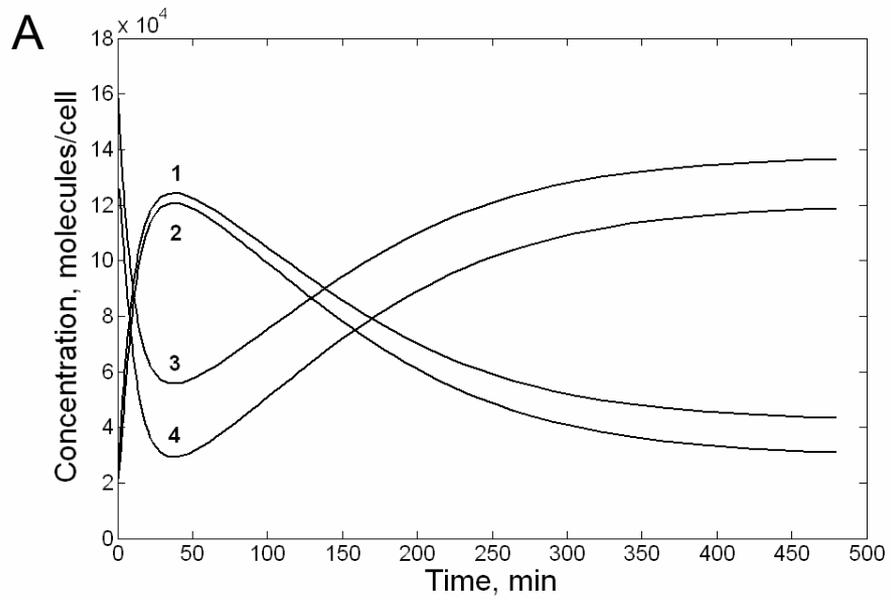

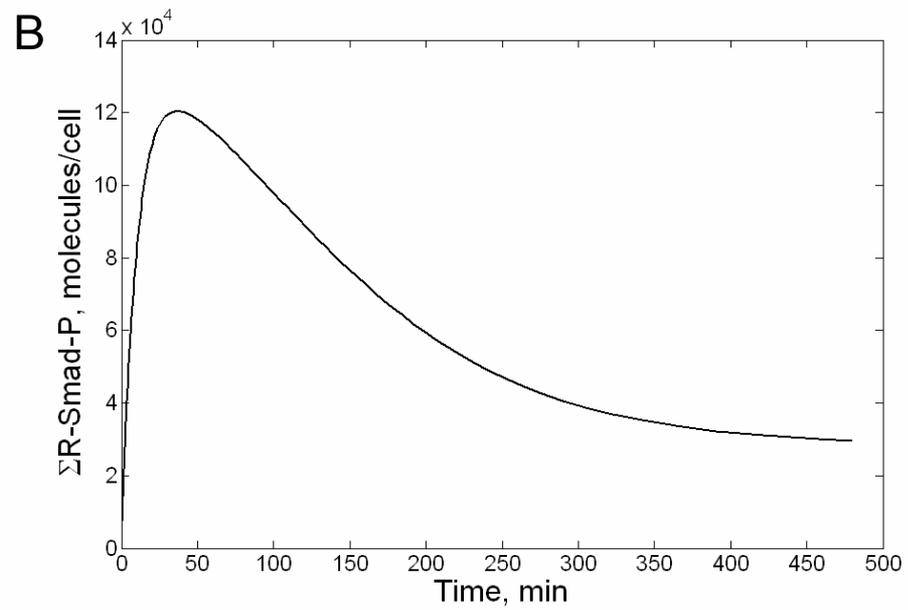

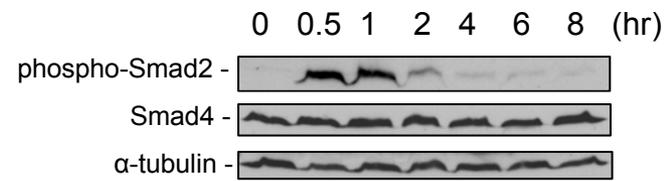

**Figure 2**.

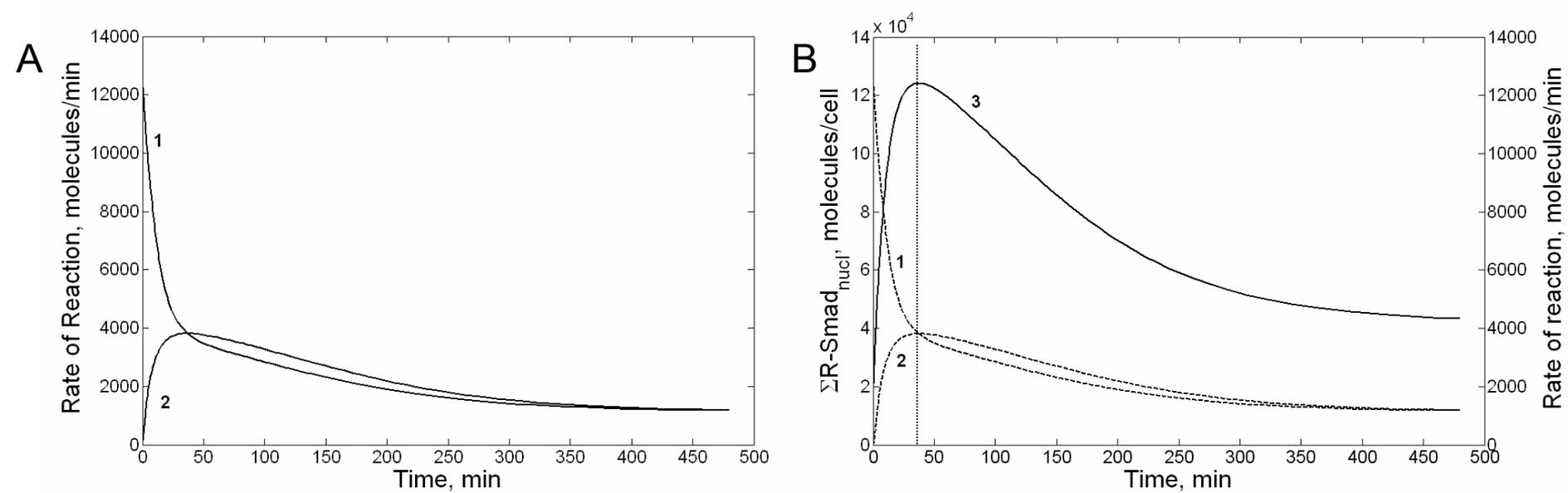

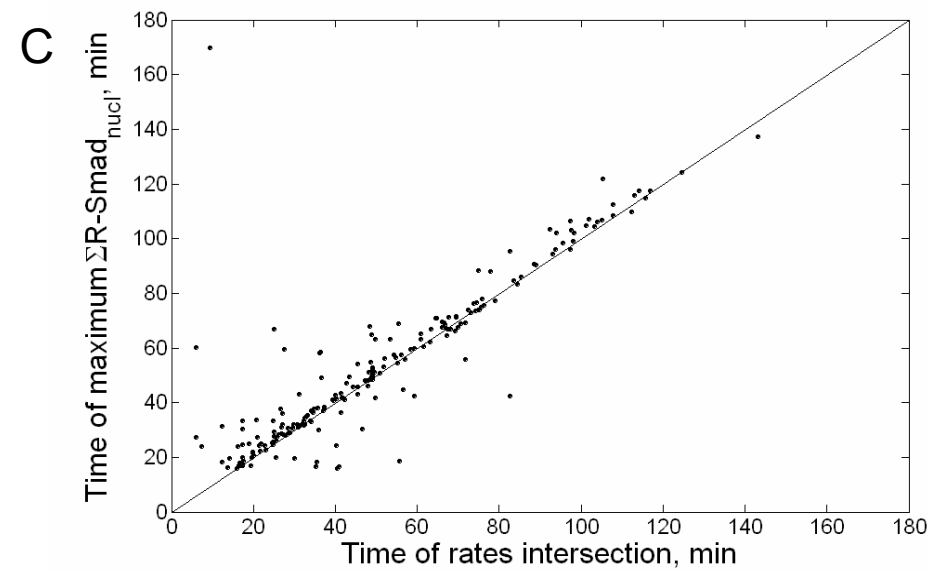

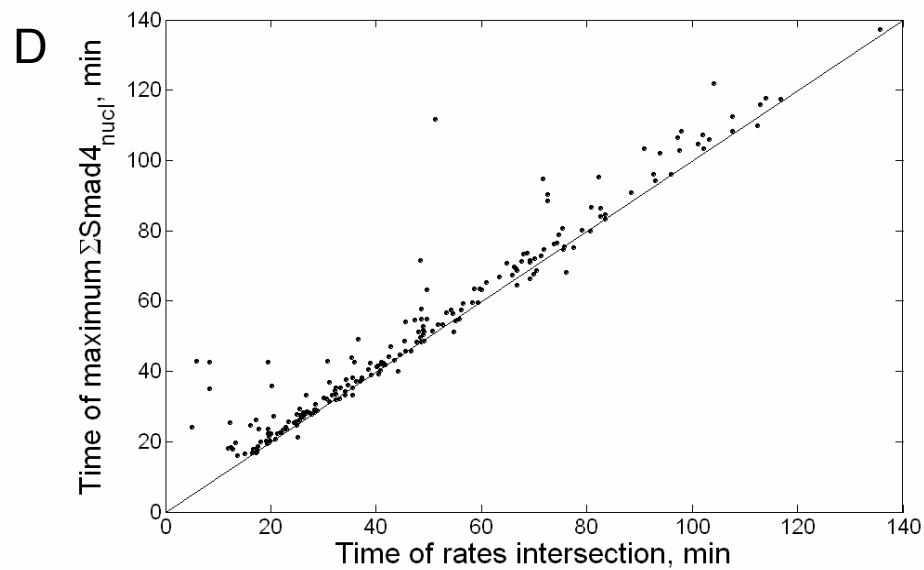

**Figure 3**.

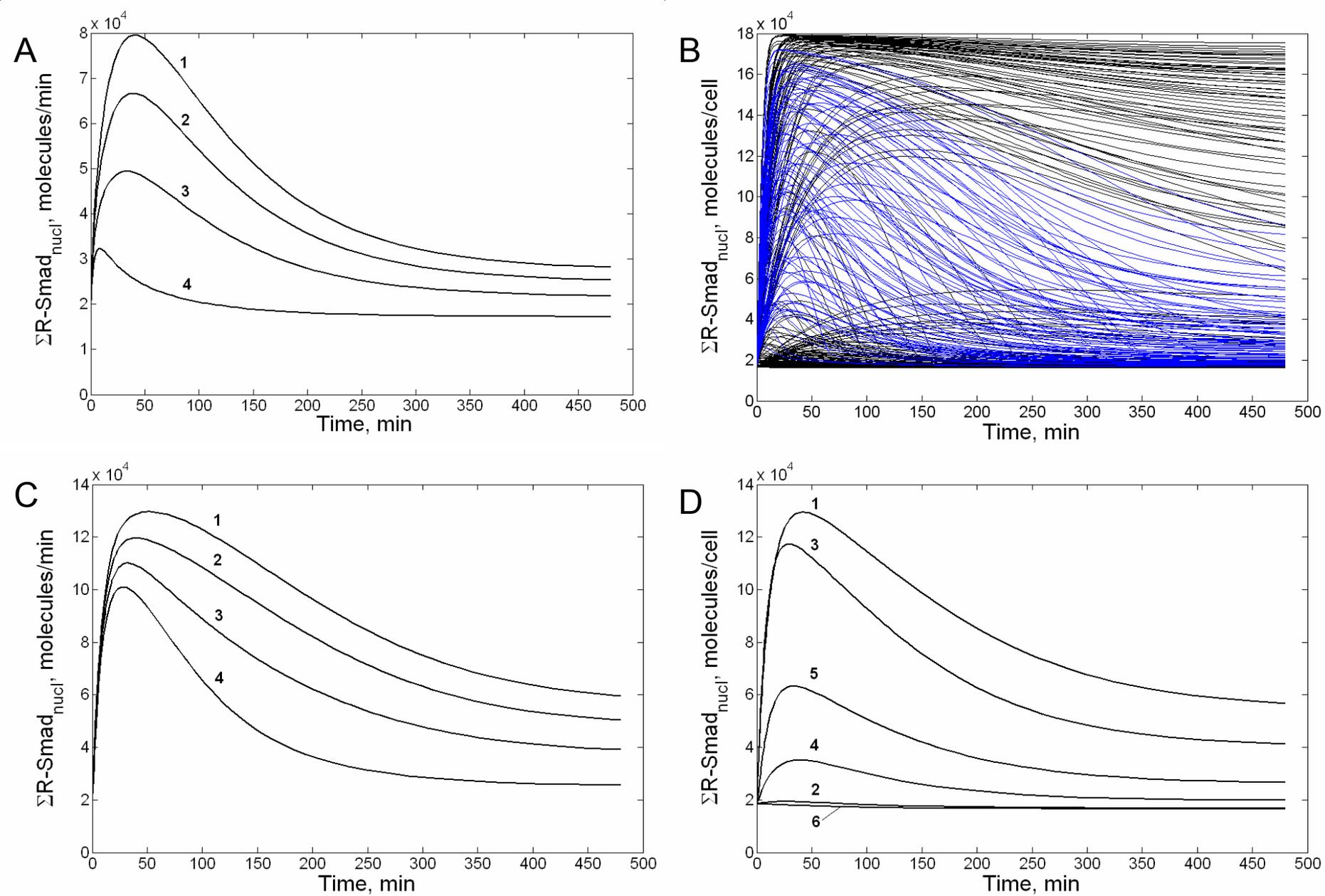

**Figure 4**.

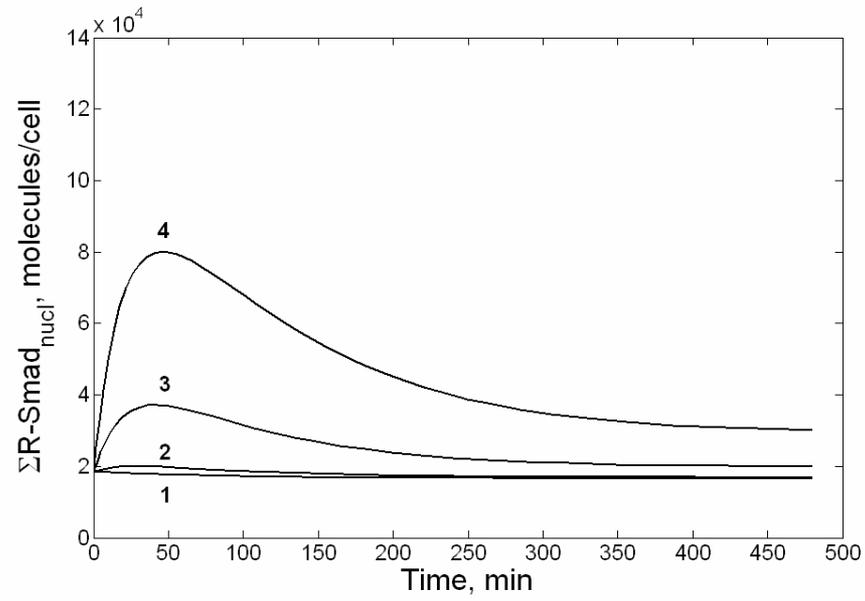

**Figure 5**.

**Tables**

**Table 1**. Model state variables and initial conditions.

| State Variable | Description | Initial Condition (molecules·cell$^{-1}$) |
|---|---|---|
| Receptors | Active receptors | $1.00 \times 10^4$ |
| R-Smad$_{cyt}$ | Cytoplasmic R-Smad | $1.62 \times 10^5$ |
| R-Smad-P$_{cyt}$ | Cytoplasmic phospho-R-Smad | 0 |
| Smad4$_{cyt}$ | Cytoplasmic Smad4 | $1.20 \times 10^5$ |
| R-Smad-P·Smad4$_{cyt}$ | Cytoplasmic phospho-R-Smad/Smad4 complex | 0 |
| R-Smad-P·Smad4$_{nucl}$ | Nuclear phospho-R-Smad/Smad4 complex | 0 |
| R-Smad$_{nucl}$ | Nuclear R-Smad | $1.80 \times 10^4$ |
| R-Smad-P$_{nucl}$ | Nuclear phospho-R-Smad | 0 |
| Smad4$_{nucl}$ | Nuclear Smad4 | $3.00 \times 10^4$ |

**Table 2**. Quantification of absolute Smad cellular concentrations in Mv1Lu cells.

| Smad Isoform | Molecules per cell[a] |
|---|---|
| Smad2 | $1.27\text{-}1.77 \times 10^5$ |
| Smad3 | $1.86\text{-}3.84 \times 10^4$ |
| Smad4 | $1.34\text{-}1.67 \times 10^5$ |

[a]95% confidence intervals, n=6 independent cell lysates

**Table 3**. Model parameter quartile values from parameters in dataset1 (n=203).

| Parameter | Units | Minimum | 25th Quartile | Median | 75th Quartile | Maximum | Range of $\log_{10}$ max & min values |
|---|---|---|---|---|---|---|---|
| $k_{cat1}$ – turnover number, TGF-beta type I receptor | $min^{-1}$ | 0.272 | 1.38 | 3.51 | 8.25 | $1.87 \times 10^2$ | 2.84 |
| $K_1$ – Michaelis-Menten constant, TGF-beta type I receptor | $molecules \cdot cell^{-1}$ | $6.33 \times 10^2$ | $9.14 \times 10^4$ | $2.89 \times 10^5$ | $9.78 \times 10^5$ | $2.27 \times 10^7$ | 4.55 |
| $k_{2a}$ – association rate constant, phosphorylated R-Smad & Smad4 in the cytoplasm | $cell \cdot molecules^{-1} \cdot min^{-1}$ | $7.73 \times 10^{-7}$ | $1.60 \times 10^{-5}$ | $6.50 \times 10^{-5}$ | $2.72 \times 10^{-4}$ | $1.51 \times 10^{-2}$ | 4.29 |
| $k_{2d}$ – dissociation rate constant, phosphorylated R-Smad/Smad4 in cytoplasm | $min^{-1}$ | $2.05 \times 10^{-5}$ | $9.73 \times 10^{-3}$ | $3.99 \times 10^{-2}$ | $1.98 \times 10^{-1}$ | 7.71 | 5.58 |
| $k_3$ – nuclear import rate, phosphorylated R-Smad/Smad4 | $min^{-1}$ | $4.75 \times 10^{-2}$ | 3.03 | 16.6 | 63.0 | $1.75 \times 10^3$ | 4.57 |
| $k_{4ne}$ – nuclear export rate, Smad4 | $min^{-1}$ | $8.86 \times 10^{-3}$ | $1.54 \times 10^{-1}$ | $7.83 \times 10^{-1}$ | 3.05 | 69 | 3.89 |
| $k_{4cn}$ – nuclear import rate, Smad4 | $min^{-1}$ | $1.87 \times 10^{-5}$ | $1.24 \times 10^{-3}$ | $4.97 \times 10^{-3}$ | $1.78 \times 10^{-2}$ | 0.398 | 4.33 |
| $k_{5nc}$ – nuclear export rate, R-Smad | $min^{-1}$ | $3.57 \times 10^{-2}$ | 1.12 | 5.63 | 27.0 | $2.47 \times 10^3$ | 4.84 |
| $k_{5cn}$[a] – nuclear import rate, R-Smad | $min^{-1}$ | $= 0.1 \ast k_{5nc}$ | $= 0.1 \ast k_{5nc}$ | $= 0.1 \ast k_{5nc}$ | $= 0.1 \ast k_{5nc}$ | $= 0.1 \ast k_{5nc}$ | 4.84 |
| $k_{6a}$ – association rate constant, phosphorylated R-Smad & Smad4 in the nucleus | $cell \cdot molecules^{-1} \cdot min^{-1}$ | $6.21 \times 10^{-7}$ | $2.88 \times 10^{-5}$ | $1.44 \times 10^{-4}$ | $7.43 \times 10^{-4}$ | $1.35 \times 10^{-2}$ | 4.34 |
| $k_{6d}$ – dissociation rate constant, phosphorylated R-Smad/Smad4 in the nucleus | $min^{-1}$ | $8.53 \times 10^{-3}$ | $2.77 \times 10^{-2}$ | $4.92 \times 10^{-2}$ | $1.16 \times 10^{-1}$ | 9.27 | 3.04 |
| $v_{max7}$ – maximal velocity, nuclear phosphatase | $molecules \cdot cell^{-1} \cdot min^{-1}$ | $5.04 \times 10^2$ | $4.99 \times 10^3$ | $1.71 \times 10^4$ | $6.33 \times 10^4$ | $6.77 \times 10^6$ | 4.13 |
| $K_7$ – Michaelis-Menten constant, nuclear phosphatase | $molecules \cdot cell^{-1}$ | 5.24 | $2.46 \times 10^3$ | $8.95 \times 10^3$ | $2.85 \times 10^4$ | $6.34 \times 10^5$ | 5.08 |

[a]$k_{5cn}$ was constrained to 10% of $k_{5nc}$ to ensure recovery of the R-Smad initial conditions towards the end of the signalling period.

**Table 4**. Parameter sensitivities from multiple regression analysis.

| Regression Model | Adjusted $R^2$ values (%) | |
|---|---|---|
| | Response variable: $\int\Sigma\text{R-Smad}_{nucl}$ | Response variable: $\int\Sigma\text{Smad4}_{nucl}$ |
| All terms included | 98.48 | 98.46 |
| **Parameter Removed:** | | |
| $k_{cat1}$ | 60.64 | 62.75 |
| $K_1$ | 61.31 | 63.58 |
| $k_{2a}$ | 98.39 | 98.41 |
| $k_{2d}$ | 98.51 | 98.49 |
| $k_3$ | 98.51 | 98.49 |
| $k_{4nc}$ | 92.74 | 83.01 |
| $k_{4cn}$ | 98.17 | 97.44 |
| $k_{5nc}$ | 98.49 | 98.48 |
| $k_{6a}$ | 91.12 | 90.38 |
| $k_{6d}$ | 85.96 | 85.28 |
| $v_{max7}$ | 87.50 | 93.40 |
| $K_7$ | 92.47 | 95.18 |

**Table 5**. Model parameters and Smad4 initial conditions used in the analyses for Figures 4 and 5.

| Parameter | Units | Figure 4C | Figure 4D Curve # | | | | | | Figure 5 |
|---|---|---|---|---|---|---|---|---|---|
| | | | 1[b] | 2 | 3 | 4 | 5 | 6 | |
| $k_{cat1}$ | $min^{-1}$ | 3.19 | 3.19 | 3.19 | 3.19 | 3.19 | 3.19 | 3.19 | 3.19 |
| $K_1$ | $molecules \cdot cell^{-1}$ | $2.09 \times 10^5$ | $2.09 \times 10^5$ | $2.09 \times 10^5$ | $2.09 \times 10^5$ | $2.09 \times 10^5$ | $2.09 \times 10^5$ | $2.09 \times 10^5$ | $2.09 \times 10^5$ |
| $k_{2a}$ | $cell \cdot molecules^{-1} \cdot min^{-1}$ | $6.50 \times 10^{-5}$ | $1.17 \times 10^{-6}$ | $1.17 \times 10^{-6}$ | $9.23 \times 10^{-2}$ | $9.23 \times 10^{-2}$ | $9.23 \times 10^{-5}$ | $9.23 \times 10^{-5}$ | $9.23 \times 10^{-7}$ |
| $k_{2d}$ | $min^{-1}$ | $3.99 \times 10^{-2}$ | 0.05 | 0.05 | 0.05 | 0.05 | 0.05 | 0.05 | 0.05 |
| $k_3$ | $min^{-1}$ | 16.6 | 16.6 | 16.6 | 16.6 | 16.6 | 16.6 | 16.6 | 16.6 |
| $k_{4nc}$ | $min^{-1}$ | 0.783 | 0.783 | 0.783 | 0.783 | 0 | 0.783 | 0 | 0 |
| $k_{4cn}$ | $min^{-1}$ | $4.97 \times 10^{-3}$ | 0.157 | 0.157 | 0.157 | $1 \times 10^4$ | 0.157 | $1 \times 10^4$ | $1 \times 10^4$ |
| $k_{5nc}$ | $min^{-1}$ | 5.63 | 5.63 | 5.63 | 5.63 | 5.63 | 5.63 | 5.63 | 5.63 |
| $k_{5cn}$ | $min^{-1}$ | 0.563 | 0.563 | 0.563 | 0.563 | 0.563 | 0.563 | 0.563 | 0.563 |
| $k_{6a}$ | $cell \cdot molecules^{-1} \cdot min^{-1}$ | $1.44 \times 10^{-4}$ | $1.05 \times 10^{-5}$ | $1.05 \times 10^{-5}$ | 0.831 | 0.831 | $8.31 \times 10^{-4}$ | $8.31 \times 10^{-4}$ | $8.31 \times 10^{-6}$ |
| $k_{6d}$ | $min^{-1}$ | $4.92 \times 10^{-2}$ | 0.05 | 0.05 | 0.05 | 0.05 | 0.05 | 0.05 | 0.05 |
| $v_{max7}$ | $molecules \cdot cell^{-1} \cdot min^{-1}$ | $1.18 \times 10^4$ | $1.18 \times 10^4$ | $1 \times 10^6$ | $1 \times 10^6$ | $1 \times 10^6$ | $1 \times 10^6$ | $1 \times 10^6$ | $1 \times 10^6$ |
| $K_7$ | $molecules \cdot cell^{-1}$ | $7.29 \times 10^4$ | $7.29 \times 10^4$ | 1 | 1 | 1 | 1 | 1 | 1 |
| $k_8$ | $min^{-1}$ | 16.6 | 16.6 | 16.6 | 16.6 | 16.6 | 16.6 | 16.6 | 16.6 |
| $K_d{}^a$ (R-Smad-P/ Smad4): | nM | | 79 | 79 | $1 \times 10^{-3}$ | $1 \times 10^{-3}$ | 1 | 1 | 100 |
| Smad4 initial conditions: | | | | | | | | | |
| $Smad4_{cyt}$ | $molecules \cdot cell^{-1}$ | $1.2 \times 10^5$ | $1.2 \times 10^5$ | $1.2 \times 10^5$ | $1.2 \times 10^5$ | 0 | $1.2 \times 10^5$ | 0 | 0 |
| $Smad4_{nucl}$ | $molecules \cdot cell^{-1}$ | $3 \times 10^4$ | $3 \times 10^4$ | $3 \times 10^4$ | $3 \times 10^4$ | $1.5 \times 10^5$ | $3 \times 10^4$ | $1.5 \times 10^5$ | varied |

[a] $K_d = k_{2d}/k_{2a} = k_{6d}/k_{6a}$. Conversion from units of molecules·cell$^{-1}$ to nM assumes a volume of $9 \times 10^{-13}$ L for the cytoplasm and $1 \times 10^{-13}$ L for the nucleus.

[b] For the baseline parameter set, we constrained $k_{2a}$, $k_{6a}$, $k_{2d}$, and $k_{6d}$ to ensure that the $K_d$ (nM) in the nucleus and cytoplasm was the same. Also, we set $k_{4cn} = 0.2 \times k_{4nc}$, to reflect the Smad4 partitioning under basal conditions. These constraints were carried through to the other parameter sets where appropriate.